\RequirePackage{fix-cm}
\documentclass[smallcondensed]{svjour3}     
\smartqed  
\usepackage{graphicx}
\usepackage{caption}
\usepackage[numbers,square,sort]{natbib}
\begin{document}

\title{PLU-Net: Extraction of multi-scale feature fusion
}

\author{Weihu Song       
}


\institute{BeiHang University \at
              \email{weihusong@buaa.edu.cn}   \at
              Beijing, 100191, China
}

\date{Received: date / Accepted: date}

\maketitle

\begin{abstract}
Deep learning algorithms have achieved remarkable results in medical image segmentation in recent years. These networks are unable to handle with image boundaries and details with enormous parameters, resulting in poor segmentation results. To address the issue, we develop atrous spatial pyramid pooling (ASPP) and combine it with the Squeeze-and-Excitation block (SE block), as well as present the PS module, which employs a broader and multi-scale receptive field at the network's bottom to obtain more detailed semantic information. We also propose the Local Guided block (LG block) and also its combination with the SE block to form the LS block, which can obtain more abundant local features in the feature map, so that more edge information can be retained in each down sampling process, thereby improving the performance of boundary segmentation. We propose PLU-Net and integrate our PS module and LS block into U-Net. We put our PLU-Net to the test on three benchmark datasets, and the results show that by fewer parameters and FLOPs, it outperforms on medical semantic segmentation tasks.
\end{abstract}

\section{Introduction}\label{sec1}
The significance of image analysis is rising in parallel with the successful application of imaging in clinical medicine. Image segmentation is a key image analysis technology which plays an essential role in imaging medicine. Deep learning technology, mainly based on deep convolutional neural network (DCNN), have solved various semantic segmentation difficulties of medical images in recent times.Although the performance of subsequent improved methods based on the U-Net has improved, some issues have emerged, including such increasing the parameters and FLOPs, and the network for image segmentation boundary and details is not good enough.In this article, we propose PLU-Net, a simple and effective network model which utilizes U-Net as a baseline to solve the problems. To start, the LS block is employed to obtain rich local information in order to improve boundary segmentation performance. Second, with its broad and multi-scale receptive field, the PS module is added to the bottom of the network to collect richer detail information. The combination of the two modules allows for excellent image boundary and detail information acquisition. Finally, the network depth is reduced by one layer, greatly increasing the model's interface speed.

\section{Related work}\label{sec2}
Other CNN models appeared in the years after the ILSVRC~\cite{ILSVRC15} competition began in 2012, including ALexNet~\cite{krizhevsky2012imagenet}, VGG~\cite{simonyan2014very}, GoogLeNet~\cite{szegedy2015going}, ResNet~\cite{he2016deep}, and SENet~\cite{hu2018squeeze}. These models are mostly utilized in image classification tasks at the image level, and many fields require more detailed image classification. This is especially true in medical imaging, where precision and speed are more important than other fields. 
\par
\textbf{deep convolutional networks}: In 2015, the Fully Convolutional Network (FCN)~\cite{long2015fully} replaced the fully connected layer with the convolution layers to output spatial mapping, allowing the model to handle images of varying sizes and considerably boosting segmentation accuracy over traditional methods. However, it still has significant flaws, such as the model's poor recognition efficiency in particular cases and the omission of global context information. At this time, U-Net was born. It uses a completely symmetrical model structure and uses an altogether new feature fusion technique than FCN: concat. Meanwhile, it reduces the size of the model and delivers excellent results with little training data, which is essential for medical segmentation. More semantic segmentation models employ U-Net as the basis for improvement because of its superior performance. U-Net++~\cite{2020UNet} improves accuracy by adding deep supervision to each layer's sub-network and better capturing some feature information lost in down-sampling and up-sampling operations.
\par
\textbf{multi-scale feature extraction}: PSPNet\cite{2016Pyramid} proposes to use the pyramid pooling module to aggregate the context information of different regions, so as to improve the extraction ability of feature information. DeepLab\cite{chen2014semantic} uses ASPP to aggregate more convolution kernels of different scales to improve the multi-scale feature extraction ability.Res-UNet~\cite{xiao2018weighted} and Dense-UNet~\cite{huang2017densely}, respectively, incorporate ResNet and DenseNet concepts into the U-Net, ResNet's residual block and DenseNet's dense block are used to effectively reduce feature information loss during transmission.
\par
\textbf{attention modules}: For each up-sampling, Attention U-Net~\cite{oktay2018attention} inserts the attention gate into U-Net, which eliminates feature redundancy caused by the repetitive employment of low-level features in multiple convolution processing.  R2U-Net~\cite{alom2018recurrent} combines the RNN and ResNet structures into a U-Net structure, allowing the structure to gain more characteristic information after each convolution.  Additionally, because transformer has a global receptive field and can acquire feature information from all pixels in an image, numerous recent works~\cite{fan2022sunet}~\cite{sha2021transformer}~\cite{chen2021transunet}~\cite{petit2021u}~\cite{lin2021ds} have merged transformer and U-Net in various ways. The models performance has improved to some level, but it has also introduced a slew of new issues. On the one hand, a transformer structure will dramatically increase the size of the model, stifling inference speed and necessitating higher hardware needs. On the other hand, it frequently requires the blessing of the pre-trained model, therefore a solid pre-trained model is critical to the model's performance. To summarize, the model's accuracy can be enhanced by learning additional feature information or lowering feature information loss during the feature map calculation. In addition, during the model design process, the long-running time induced by the growth of the model size must be taken into account. 
\begin{figure*}[htb]
	\begin{minipage}[b]{0.48\linewidth} 
		\centering
		\includegraphics[width=5cm]{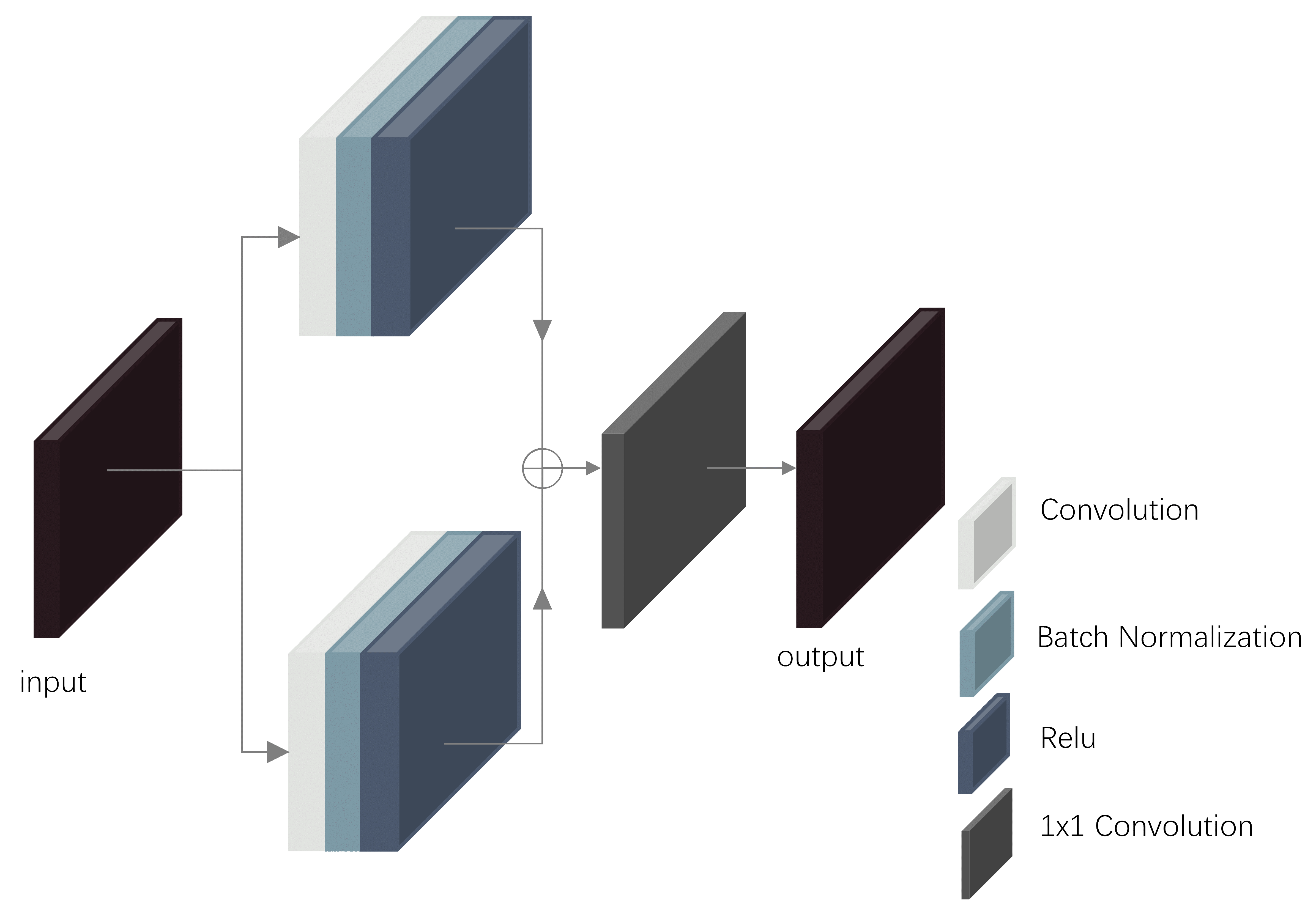} 
		\centerline{(a) LG block} 
		\label{lg} 
	\end{minipage}
	\begin{minipage}[b]{0.48\linewidth}
		\centering
		\includegraphics[width=5cm]{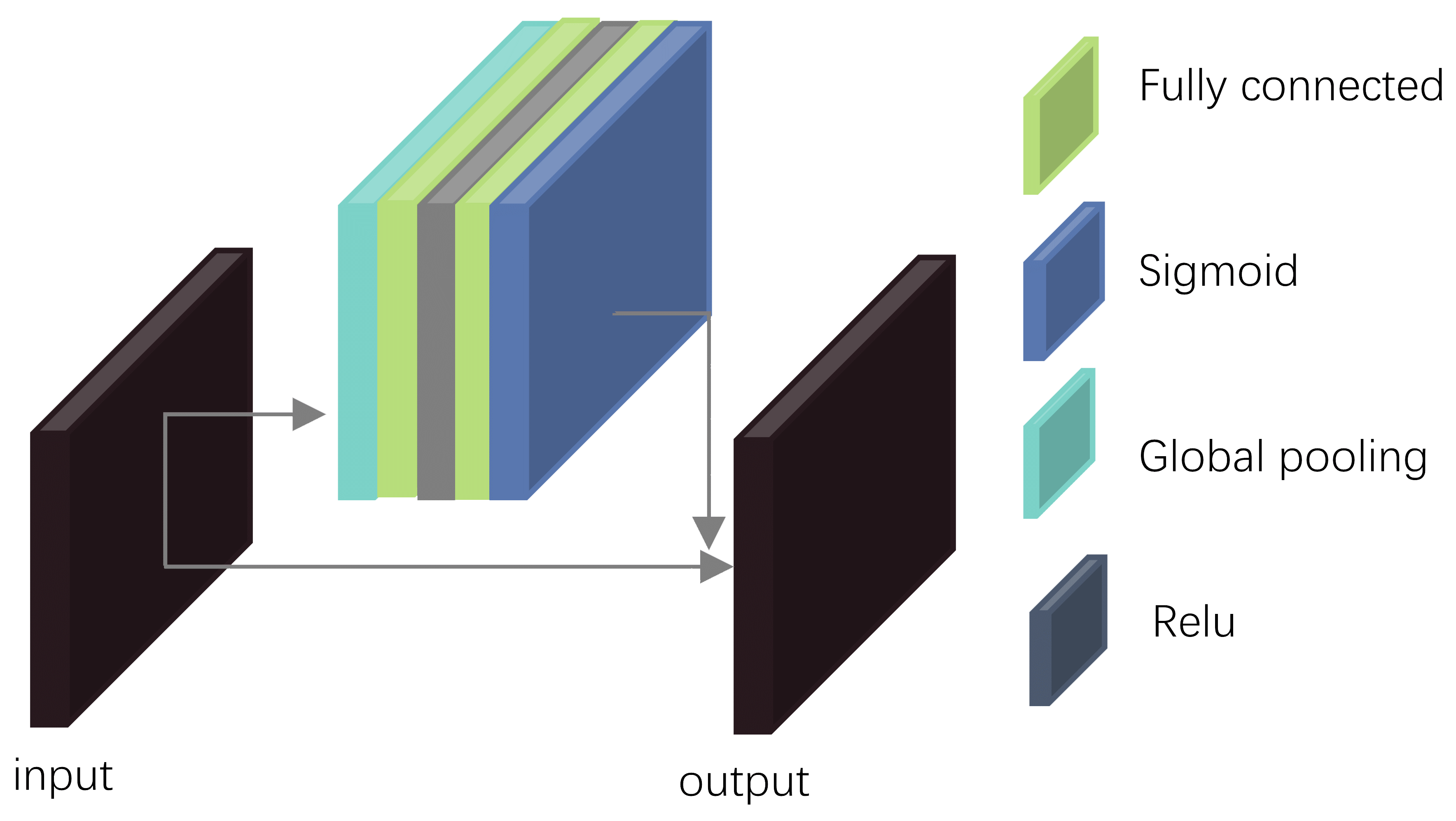}
		\centerline{(b) SE block} 
		\label{se} 
	\end{minipage}
	\begin{minipage}[b]{0.48\linewidth}
		\centering
		\includegraphics[width=10cm]{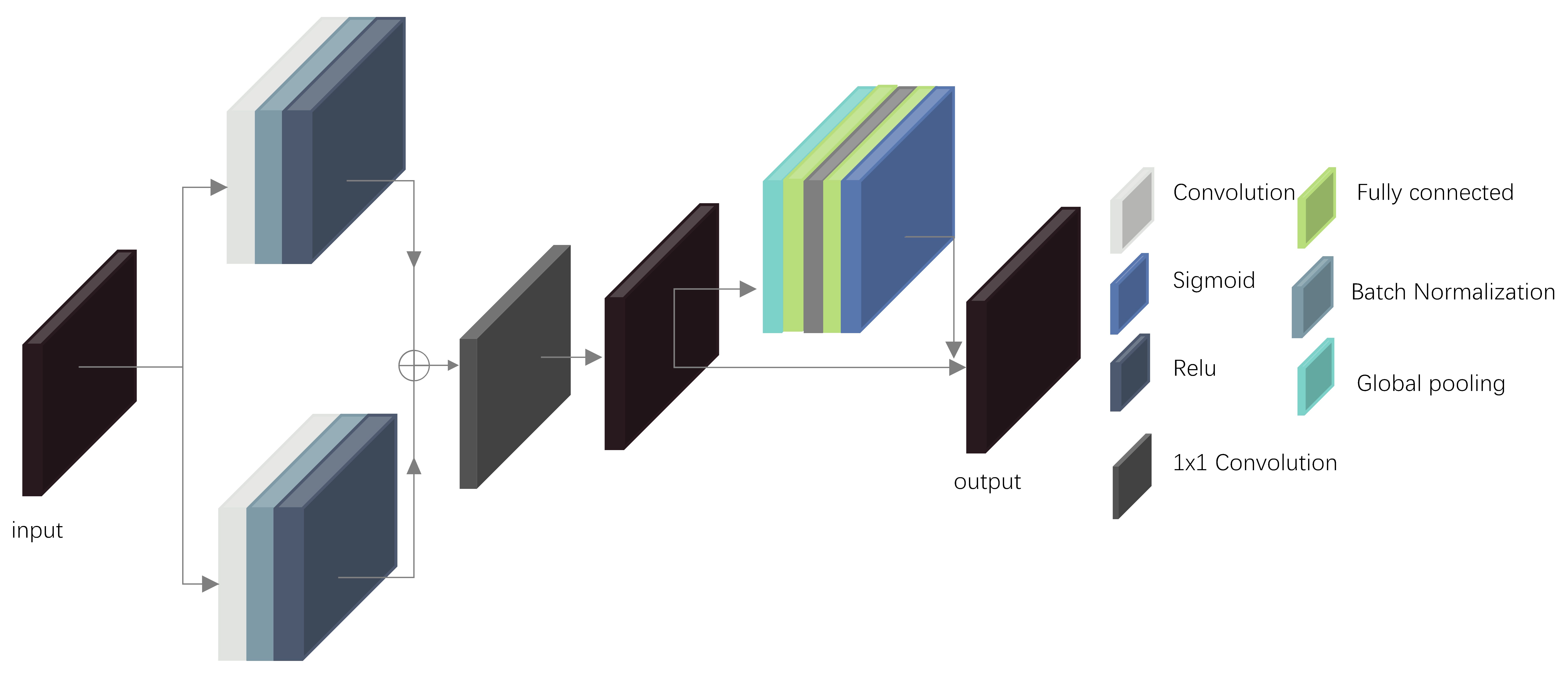}
		\centerline{(c) LS block} 
		\label{ls} 
	\end{minipage}
	\caption{Comparison of three different blocks.}
	\label{fig:Comparison1}
\end{figure*}

\section{Methodology}\label{sec3}
\subsection{LS block}\label{subsec31}
The Conv block in the original U-Net network consists of two 3x3 convolution operations, two batch normalization operations, and two nonlinear activations (ReLU). However, we notice that this structure has a loss of local information, so we propose the local guided block (LG block, shown as Fig.\ref{fig:Comparison1}), which is divided into two branches and is made up of two 3x3 dilated convolution operations with dilated rates equal to 1 and 3, The results of two dilated convolution operations are then concatenated to enhance feature propagation. Then, to achieve cross-channel information fusion, a 1x1 convolution operation is employed, and nonlinear features are added on the assumption that the size of the feature map remains unchanged. To achieve channel information adaptation, we inserted SE block after LG block to make LS block (shown as Fig.\ref{fig:Comparison1}), similar to PS module. LG block, as comparison to the original convolution block, minimizes the amount of calculation while obtaining richer feature information with a large receptive field thanks to the addition of double branch structure and dilated convolution. It further realizes the adaptation of channel feature information by adding SE block.

\subsection{PS module}\label{subsec32}
\begin{figure*}[htb]
	\begin{minipage}[b]{0.48\linewidth}
		\centering
		\includegraphics[width=5cm]{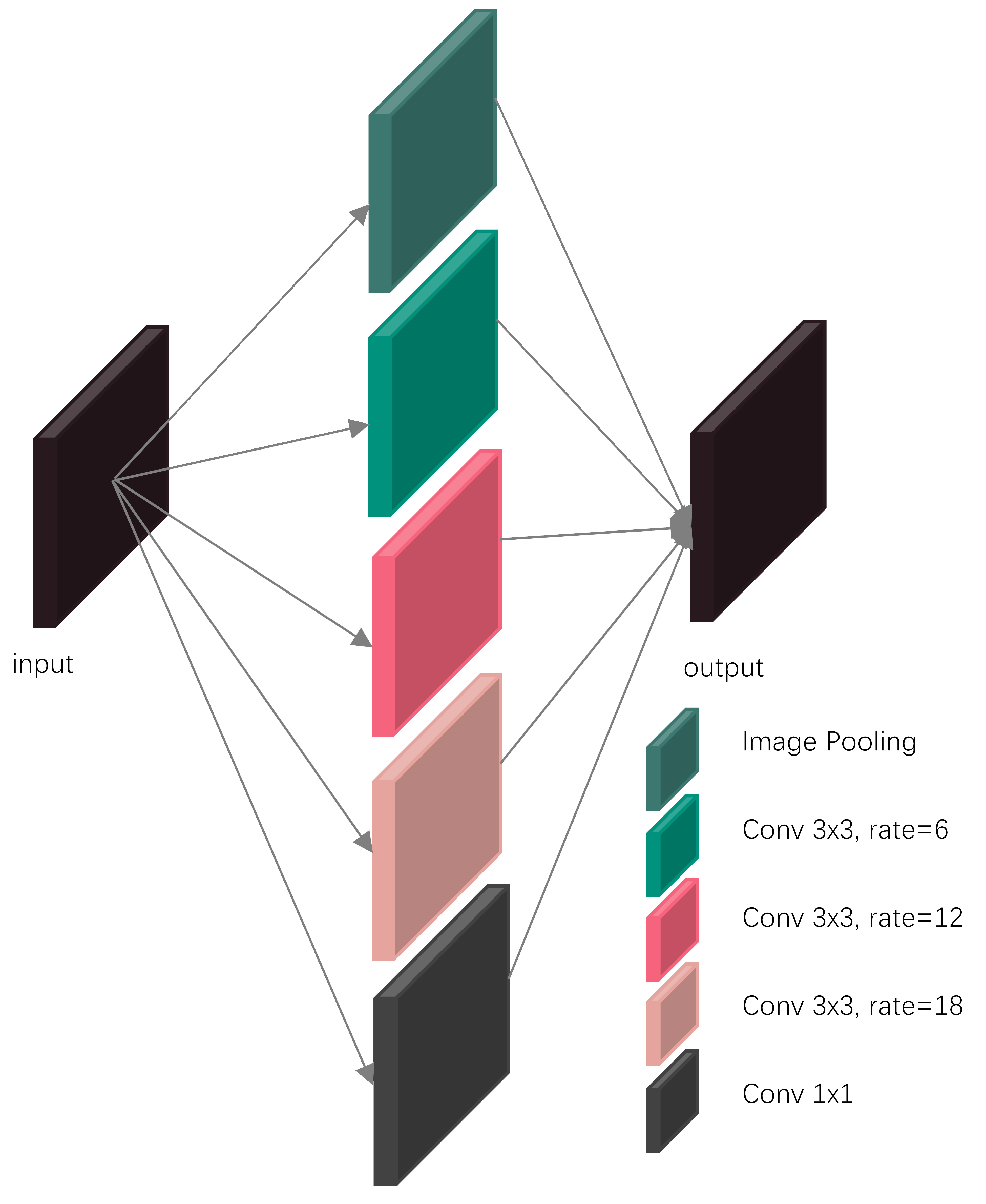}
		\centerline{(a) ASPP}
		\label{aspp}
	\end{minipage}
	\begin{minipage}[b]{0.48\linewidth}
		\centering
		\includegraphics[width=5cm]{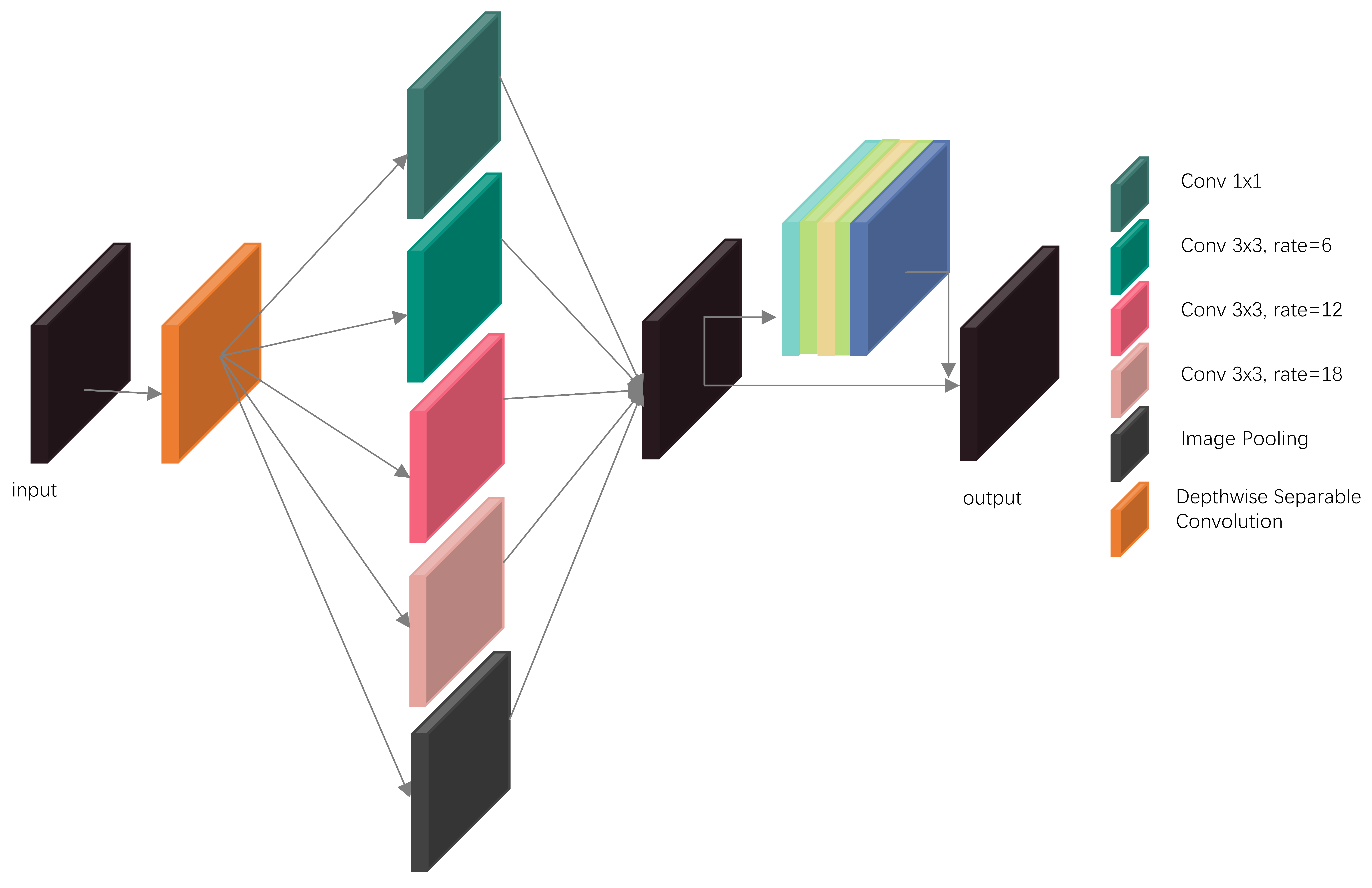}
		\centerline{(b) PS module}
		\label{saspp}
	\end{minipage}
	\caption{Comparison of different feature extractors.}
	\label{fig:Comparison2}
\end{figure*}

ASPP first was proposed in DeepLabv1~\cite{chen2014semantic} and then improved in DeepLabv2~\cite{chen2017deeplab} and DeepLabv3~\cite{chen2017rethinking}, as seen in Fig.\ref{fig:Comparison2}. Deeplabv3's ASPP, which comprises of one 1x1 convolution, three dilated convolutions, and one global average pooling operation, is used as the foundation. The employment of the global average pooling method in up-sampling produces duplicate information and degrades prediction performance, according to experimental verification. As a result, we remove the global average pooling and replace the ordinary void convolution with the depthwise separable convolution~\cite{sifre2014rigid}, resulting in a reduction of roughly five times the number of parameters compared to the original ASPP. The ASPP's multi-scale structure allows it to gather more feature information and utilise larger receptive fields, however more feature information may contain duplicated data, lowering the performance of the system.To alleviate the impact of redundant information, we choose to employ the SE block in SENet~\cite{hu2018squeeze} to increase the weight of important channel information while decreasing the weight of worthless channel information. The SE block uses squeeze and excitation two processes to learn the importance of each channel's features, and then strengthens the relevant channel while weakening the idle channel to achieve adaptive feature channel calibration. As a result, a PS module is proposed (shown as Fig.\ref{fig:Comparison2}). After getting additional feature information and employing the large receptive field, this module may combine the advantages of the ASPP module and the SE block to suppress redundant information and strengthen the important channel feature information.

\subsection{Network Architecture}\label{subsec33}
\begin{figure}[htb]
	\centering
	\includegraphics[width=\textwidth]{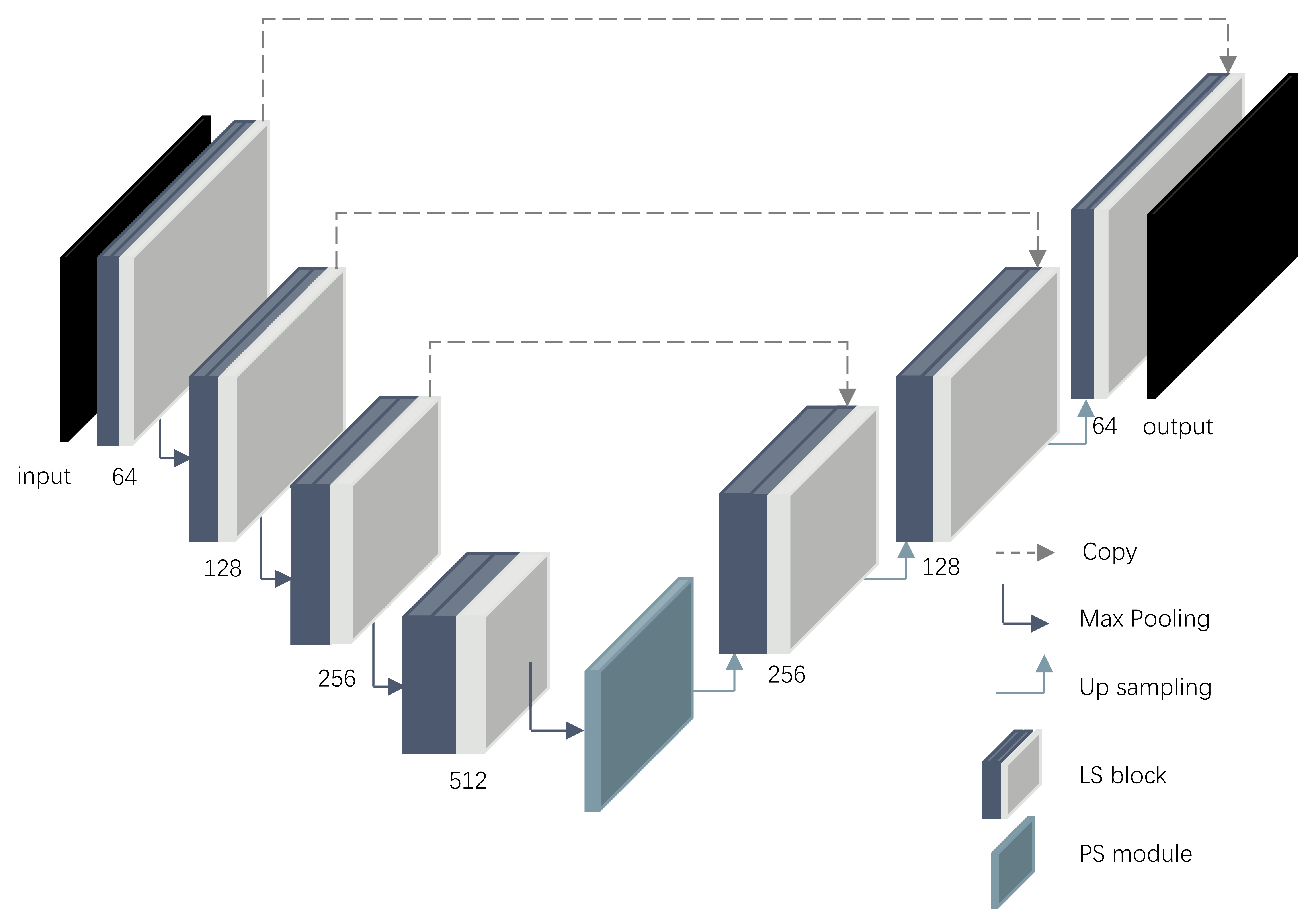}
	\caption{Proposed PLU-Net architecture.}
	\label{psu}
\end{figure}

Our PLU-Net, as shown in Fig.\ref{psu}, improves on the original U-Net network architecture by substituting the LS block for the convolution block in the down-sampling and up-sampling pathways. The LS block's double-branch structure effectively ensures that information loss is minimized at each layer of operation, while feature reuse and successful propagation are ensured by the bigger receptive field. In addition, the U-Net network's up-sampling and down-sampling periods were reduced from four to three, and a PS module was added at the end of the down-sampling. We may considerably minimize the calculation amount and make the model more lightweight by reducing the number of down-sampling and up-sampling channels in the last layer of the U-Net network, which is 1024. We employ four branches and a greater dilation rate at the same time to obtain a bigger receptive field and consequently richer feature information. The up-sampling step that follows can be completed efficiently. Our network structure can now achieve better performance with fewer parameters and FlOPs because to the combination of these enhancements.

\begin{figure*}[h]
	\begin{minipage}[b]{.3\linewidth}
		\centering
		\includegraphics[width=1.5cm]{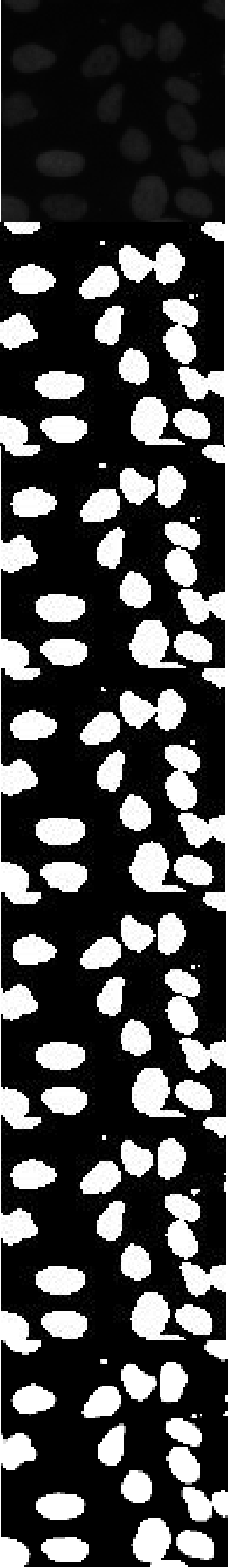}
		\centerline{(a) DSB2018}
	\end{minipage}
	\begin{minipage}[b]{0.3\linewidth}
		\centering
		\includegraphics[width=2cm]{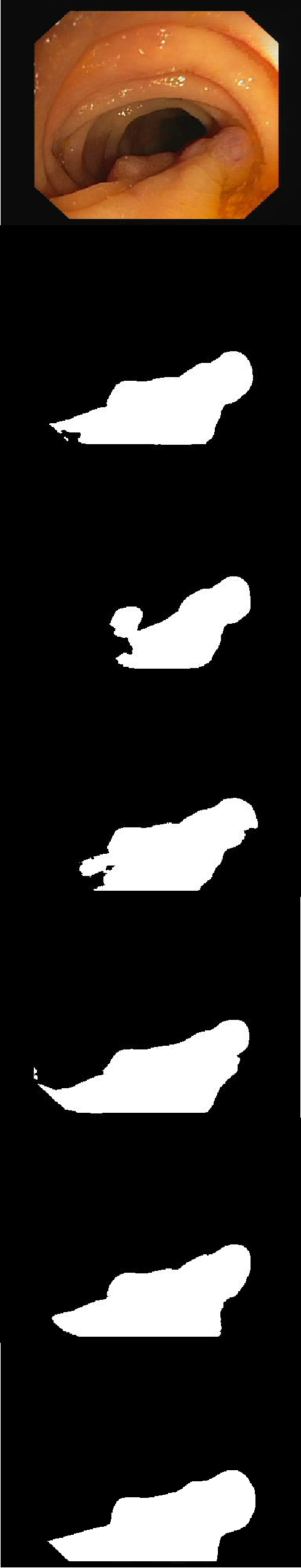}
		\centerline{(b) CVC}
	\end{minipage}
	\begin{minipage}[b]{0.3\linewidth}
		\centering
		\includegraphics[width=1.5cm]{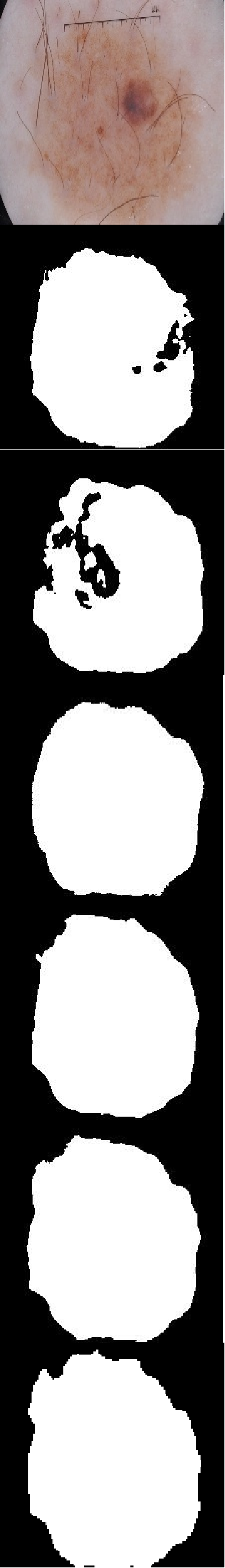}
		\centerline{(c) ISIC2018}
	\end{minipage}
	\caption{Qualitative comparison of segmentation results for nuclei, colon, and skin lesion datasets, from top to bottom are Image, U-Net, U-Net++, MultiResUnet, DoubleUNet, PLU-Net, Ground Truth}
	\label{fig:QC}
\end{figure*}

\section{Experiments and Results}\label{sec4}

\subsection{Datasets}\label{subsec41}
Because both the PS module and the LS block are modular, they can simply be utilized to substitute convolution processes in various network architectures. We designed three models, PU-Net (conv+PS module), LUNet(LS block), and PLU-Net (LS block+PS module), to demonstrate the robustness of our model, in addition to the original U-Net (conv+null here A+B means A for down-sampling and up-sampling pathway and B for PS module, the same below.) We evaluated the models on three biomedical image segmentation datasets in the study.

\subsubsection{Polyp Segmentation}\label{subsubsec42}
CVC-ClinicDB\cite{Bernal2015WMDOVAMF}(CVC for short) is from colonoscopy videos and contains 612 polyp images. 
We use the original size
384x288 of image and is randomly split into train set ($60\%$), validation set ($20\%$), and test set ($20\%$). Also, we scale the original images equally (resize it from $512 \times 512$ to $256 \times 256$).

\subsubsection{Nuclei Segmentation}\label{subsubsec43}
In most cancer grading schemes, nuclei segmentation has far-reaching significance because nuclear morphology is one of the important components. The dataset is derived from Kaggle 2018 Data Science Bowl\footnote{https://www.kaggle.com/c/data-science-bowl-2018/data} (DSB2018 for short). It contains 670 nucleus images and is randomly split into train set ($60\%$), validation set ($20\%$), and test set ($20\%$). Also, we resize the original images to $96 \times 96$.

\subsubsection{Skin Lesion Segmentation}\label{subsubsec44}
Computer-aided automatic diagnosis of skin cancer is an inevitable trend, and skin lesions segmentation as the first step is urgent. The dataset is from MICCAI 2018 Workshop - ISIC2018: Skin Lesion Analysis Towards Melanoma Detection~\cite{codella2019skin}~\cite{tschandl2018ham10000} (ISIC2018 for short). It contains 2594 images and is randomly split into train set ($60\%$), validation set ($20\%$), and test set ($20\%$). For better model training and result display, we resize all the original images to $224 \times 224$.

\subsection{Experimental Setup}\label{subsec45}
We use three datasets to compare the U-Net, PU-Net, LU-Net, U-Net++, MultiResUnet\cite{ibtehaz2020multiresunet}, DoubleUNet\cite{9183321}, and PLU-Net architectures. We chose U-Net because of its widespread use and relevance in medical image segmentation, as well as the fact that it serves as the foundation for numerous network architectures. The kernel size is set to $3 \times 3$ with dilation values of 1 and 3 correspondingly in the LS block, followed by batch normalization and ReLU. Furthermore, the PS module employs depth-wise separable convolution, the results of which are fed into four atrous convolutions with kernel sizes of $3 \times 3$ and dilation values of 1, 6, 12, and 18 respectively. The output size is determined by concatenating the results of four atrous convolutions using 1x1 convolution. For the DSB2018 dataset, we used a batch size of 16, four for the ISIC2018 dataset, and two for the CVC dataset. The optimizer is Adam~\cite{kingma2014adam}, and the two momentum terms are $0.5$ and $0.999$, with a learning rate of 0.0003. The epoch is set to 100, and the loss function is Binary CrossEntropy Loss(BCELoss). All of the experiments are run on four NVIDIA TITAN V GPUs with 12GB of RAM each, using PyTroch~\cite{NEURIPS2019_9015}.
\begin{table}[]
	\caption{Evaluation of proposed PLU-Net}
	\resizebox{\linewidth}{!}{
	\begin{tabular}{cccccccc}
		\hline
		Dataset  & Methods         & PC     & SE     & F1              & JS              & Params(M)     & FLOPs(G)       \\ \hline
		DSB2018  & U-Net\cite{ronneberger2015u}           & 0.8965 & 0.9064 & 0.9014          & 0.8205          & 34.53         & 9.21           \\ 
		& U-Net++\cite{2020UNet}         & 0.8892 & 0.9184 & 0.9036          & 0.8237          & 36.62         & 19.41          \\
		& MultiResUnet\cite{ibtehaz2020multiresunet} & 0.9432 & 0.8401 & 0.8887          & 0.7977          & 7.24         & 2.11           \\
		& DoubleUNet\cite{9183321}      & 0.8808 & 0.9298 & 0.9046          & 0.8249          & 18.84         & 6.21           \\
		& LU-Net          & 0.9067 & 0.9015 & 0.9040          & 0.8258          & 29.29         & 6.79           \\
		& PU-Net          & 0.8912 & 0.9157 & 0.9032          & 0.8234          & 38.19         & 9.32           \\
		& PLU-Net         & 0.9025 & 0.9099 & \textbf{0.9062} & \textbf{0.8279} & \textbf{6.22} & \textbf{4.99}  \\ \hline
		CVC     & U-Net\cite{ronneberger2015u}           & 0.8001 & 0.9087 & 0.8509          & 0.7385          & 34.53         & 110.49           \\ 
		& U-Net++\cite{2020UNet}         & 0.7973 & 0.9632 & 0.8724          & 0.7706          & 36.62         & 232.92          \\
		& MultiResUnet\cite{ibtehaz2020multiresunet} & 0.7929 & 0.9495 & 0.8641          & 0.7562          & 7.24         & 25.3           \\
		& DoubleUNet\cite{9183321}      & 0.8637 & 0.9222 & 0.8920          & 0.8249          & 18.84         & 74.52           \\
		& LU-Net          & 0.8591 & 0.9351 & 0.8954          & 0.8102          & 29.29         & 81.42           \\
		& PU-Net          & 0.8727 & 0.9096 & 0.8807          & 0.7979          & 38.19         & 111.85           \\
		& PLU-Net         & 0.9139 & 0.8832 & \textbf{0.8983} & \textbf{0.8125} & \textbf{6.22} & \textbf{59.9}  \\ \hline
		ISIC2018 & U-Net\cite{ronneberger2015u}           & 0.8449 & 0.9038 & 0.8734          & 0.7665          & 34.53         & 50.13           \\ 
		& U-Net++\cite{2020UNet}         & 0.8342 & 0.9156 & 0.8730          & 0.7688          & 36.62         & 105.68          \\
		& MultiResUnet\cite{ibtehaz2020multiresunet} & 0.8223 & 0.9340 & 0.8746          & 0.7732          & 7.24         & 11.48           \\
		& DoubleUNet\cite{9183321}      & 0.8567 & 0.9007 & 0.8781          & 0.7779          & 18.84         & 33.18           \\
		& LU-Net          & 0.8678 & 0.8933 & 0.8804          & 0.7802          & 29.29         & 36.94           \\
		& PU-Net          & 0.8556 & 0.8981 & 0.8763          & 0.7771          & 38.19         & 50.75           \\
		& PLU-Net         & 0.8774 & 0.9152 & \textbf{0.8959} & \textbf{0.8061} & \textbf{6.22} & \textbf{27.18}  \\ \hline
	\end{tabular}}
    \label{table1}
\end{table}

\subsection{Result and Discussion}\label{subsec46}
To better show the experimental results, we considered several performance metrics, including Precision (PC, Eq.\ref{con:ac}), Sensitivity (SE, Eq.\ref{con:se}), F1-score(F1, which is also known as Dice coefficient, DC, Eq.\ref{con:f1}) and Jaccard similarity (JS, Eq.\ref{con:js}). Variables involved in these formulas are True Positive (TP), False Positive (FP), True Negative (TN), False Negative (FN), Ground Truth (GT), and Segmentation Result (SR).

\begin{equation}
	PC=\frac{TP}{TP+FP}
	\label{con:ac}
\end{equation}

\begin{equation}
	SE=\frac{TP}{TP+FN}
	\label{con:se}
\end{equation}

\begin{equation}
	F1=2 \frac{SE * PC}{SE + PC}=2 \frac{|GT \cap SR|}{|GT|+|SR|}=DC 
	\label{con:f1} 
\end{equation}

\begin{equation}
	JS=\frac{|GT \cap SR|}{|GT \cup SR|}
\label{con:js}
\end{equation}
Table.\ref{table1} illustrates the results of our experiments using our proposed model and various state-of-the-art U-Net models, such as U-Net++, MultiResUnet, and DoubleUNet, while Table.\ref{table1} demonstrates the segmentation outcomes of three different biomedical image segmentation tasks. Table.\ref{table1} shows that our proposed models LU-Net, PU-Net, and PLU-Net are all superior to U-Net. F1 in JS is superior to U-Net among them. On CVC, our model outperforms U-Net by more than 6 and 8 points in F1 and JS, respectively, when compared to U-Net. LU-Net and PU-Net results, on the other hand, reveal that they are superior than U-Net in JS and F1, with PLU-Net outperforming all other models, proving the superiority of LG Block and the capability of PS module. Furthermore, the segmentation results of the three segmentation tasks in Fig.\ref{fig:Comparison2} show the model's advantages. In nucleus segmentation, our model performs better on the edges, and in Polyp Segmentation, the model presented in this paper greatly outperforms other models in terms of segmentation performance. Unlike other models with smooth boundary processing, our model has more refined boundary processing in skin lesion segmentation.

\section{Conclusion}\label{sec5}
In this paper, We propose an LS block for learning local feature information from a big reception field and a PS module for learning more deep information from a wider reception field. Furthermore, we design a lightweight network PLU-Net with fewer parameters and FLOPs, which can handle boundaries and details well for medical images, based on the Local Guided block and PS module. Experiments on colon cancer, nuclei, and skin lesion segmentation demonstrate the advantages of the suggested PLU-Net for generating high-quality segmentation results.

\section*{Conflict of interest}
The authors report no conflict of interest.

\section*{Data availability statement}
The data that support the findings of this study are available from the corresponding author upon reasonable request.
%
%


%
%

\bibliographystyle{spbasic}      
\bibliography{sn-bibliography}   


\end{document}